\documentclass[prd,preprint,a4paper,superscriptaddress,nofootinbib,11pt]{revtex4}
\usepackage{amssymb}
\usepackage{graphicx}
\usepackage{amsmath}
\usepackage{eurosym}
\usepackage{dsfont}
\usepackage{amsmath}
\usepackage{amssymb}
\usepackage{amsthm}
\usepackage{stmaryrd}
\usepackage{mathrsfs}
\usepackage{textcomp}
\usepackage{amsfonts}
\usepackage{txfonts}
\usepackage[T1]{fontenc}

\begin{document}

\title{Different realizations of $\kappa$- momentum space and relative-locality effect}

\author{S. Meljanac \footnote{
meljanac@irb.hr}}
\affiliation
{Rudjer Bo\v{s}kovi\'{c} Institute, Bijeni\v{c}ka c.54, HR-10002 Zagreb, Croatia}
\author{A. Pacho{\l } \footnote{
pachol@hi.is}}
\affiliation{Science Institute, University of
Iceland, Dunhaga 3, 107 Reykjavik, Iceland}
\author{A. Samsarov \footnote{
asamsarov@irb.hr}}
\affiliation
{Rudjer Bo\v{s}kovi\'{c} Institute, Bijeni\v{c}ka c.54, HR-10002 Zagreb, Croatia}
\author{Kumar S. Gupta \footnote{%
kumars.gupta@saha.ac.in}}
\affiliation{Theory Division, Saha Institute of Nuclear Physics, 1/AF
Bidhannagar, Calcutta 700064, India}
\begin{abstract}
We consider different realizations for momentum sector of $\kappa$-Poincare Hopf algebra, which is associated with a curved momentum space. We show that the notion of the particle mass as introduced recently by Amelino-Camelia et al. in the context of relative-locality is realization independent for a wide class of realizations, up to linear order in deformation parameter $l$. On the other hand, the time delay formula clearly shows a dependence on the choice of realization.
\end{abstract}

\maketitle

\section{Introduction}

A recently postulated idea on relative-locality \cite{rel-loc1,rel-loc2,rel-loc3} proposes to describe a 
"classical nongravitational regime", where both $\hbar $ and $G$ are negligible, but their ratio  $M_{P}\sim \sqrt{\frac{\hbar }{G}}$ provides an energy scale given by the Planck mass. The emergence of such an energy scale 
provides the motivation to consider the momentum space as the
fundamental entity and leads to the study of its geometry. Various
features of this momentum space geometry can be described by a
noncommutative algebra known as the $\kappa$-Minkowski algebra
\cite{L1,L2,L3,L4}, which is associated with a curved momentum
    space \cite{rel-loc1,rel-loc2,rel-loc3,Gubitosi:2011ej}. The framework adopted here consists of this curved momentum space together with the definition of particle mass as a geodesic distance in such a space. These ingredients lead to the notion of relative-locality, whereby events that are coincident for a pair of nearby observers may not be so when they are separated in spacetime. In addition, the $\kappa$-Minkowski algebra can be used to analyze the time-delay of signals coming from gamma ray bursts, which could be a signature of Planck scale physics \cite{Fermi,recent,ellis1,ellis2,smolin1,us}.

The noncommutative $\kappa$-Minkowski algebra and its symmetry quantum
 group is known in an infinite number of realizations in terms of
 commutative coordinates and derivatives \cite{st1,st3,mssg,mk-j,bp,bp1}.
  Each of these realizations corresponds to a certain ordering
 prescription. The intimate link between different realizations of the 
 noncommutative $\kappa$-Minkowski algebra and its symmetry quantum
 group and the corresponding orderings is elaborated in detail in
 Ref. \cite{mssg}.
 In the context of the relative-locality framework, the majority
 of the work done so far uses a particular realization of the
 $\kappa$-Poincare Hopf algebra, the so called Majid-Ruegg
 bicrossproduct realization. It is a natural question if other
 realizations  can provide further insight into the consequences of the
 relative-locality framework. It might happen that different realizations point to a universality of certain physical results. On the other hand, if certain predictions depend on the choice of the realizations, that can be used to constrain the allowed class of realizations.

In this paper, we shall work within a particular class of realizations of the $\kappa$-Minkowski algebra that is much broader than just the single Majid-Ruegg bicrossproduct realization. We shall show that the linearized mass formula obtained from the geometry of the momentum space is independent of the realizations within the chosen class. On the other hand, the time delay in the observation of two particles emitted simultaneously depends explicitly on the choice of the realization. If such time delays can be experimentally measured, that would lead to phenomenological constraints on the allowed class of realizations of the $\kappa$-Minkowski algebra.

The analysis here is based on two ingredients. The first one
    is utilising the nontrivial geometrical properties of the
    momentum space, as well as of the phase space and the second one is
    the notion of the particle mass as introduced in
    \cite{rel-loc1,rel-loc2,rel-loc3}. Both of these ingredients serve
    to define a relative-locality framework, within which we want to
    find what effect the change of realization has on certain physical features, such as the photon time delay
   and the velocities of massive and massless particles.

 Thus we start this paper with the general $\kappa$-Poincare momenta
    realization which is used to obtain a general form of the metric
    on momentum space. Explicit calculations of the Christoffel
    symbols and geodesic equation in Sec.III are provided for this
    certain class of realizations, which is still much broader than
    the class previously considered in the literature. Section IV
    starts with the deformed Poisson brackets which via the Hamilton
    equations provide the solutions for the particle worldlines. These
    particle worldlines appear to explicitly depend on the realization.
   Here we find that for the observer situated at the detector, the
    two simultaneously emitted photons of different energies will
    arrive at the detector with
    some time difference, exhibitting  the time delay in arrival times
    for the two particles. This time delay
    is found to depend on the choice of realization. The velocity of
    the massive particle is also found to depend on the choice of
    realization, while  interestingly, the velocity of  massless particles 
    was not found to be realization sensitive. Concluding remarks close this paper.

\section{Generalized Metric}

$\kappa $-Poincare inspired picture can be used as one of the illustrations
of curved momentum-space geometry (as well as providing an example of the
energy-momentum sector of DSR theory). In \cite{rel-loc3} it was shown that by
using the so-called Majid-Ruegg (bicrossproduct) realization for momenta one
gets that the connection (parallel transport) is nonmetric and
torsion-full. However, one is not limited to this one basis of the $\kappa $%
-Poincare momenta sector and it is possible to consider the more general
realization for the momenta, which can be written as \cite{mssg},\cite{mk-j},\cite{bp}:
\begin{equation}
P_{i}=\frac{p_{i}}{\varphi \left( A\right) }Z^{-1},\qquad P_{0}=\frac{%
Z^{-1}-Z}{2l}+\frac{l}{2}Z^{-1}\frac{p_{i}{}^{2}}{\varphi ^{2}\left(
A\right) },\qquad P_{4}=\frac{Z+Z^{-1}}{2l}-\frac{l}{2}Z^{-1}\frac{%
p_{i}{}^{2}}{\varphi ^{2}\left( A\right) }  \label{gen_real_P}
\end{equation}%
for any $\psi ,\varphi$. 
In the following we use the Lorentzian metric $\eta _{\mu \nu }=(+,-,-,-)$ and the notation: $A=ia\cdot \partial =-a\cdot p$. We also choose $a=(l,0,0,0)$ and in the quantities like $p_i^2=p_i p_i$ ($i=1,2,3$) the summation over space indices is understood. Also in the above realizations we used $Z=e^{\Psi \left( A\right) }$ with $\Psi \left( A\right) =\int_{0}^{A}\frac{dt}{\psi \left( t\right) }$\footnote{The functions $\psi,\varphi$ are related to different realizations of $\kappa$-Minkowski spacetime and will be discussed in Section IV.}, where $Z$ is the so-called shift operator which
satisfies $\left[ Z,p_{\mu }\right] =0$.

Such coordinates $P_{I}=\left( P_{\mu },P_{4}\right) $ (\ref{gen_real_P})
satisfy the (hyperboloid) condition \cite{jkg-now}: 
\begin{equation}\label{hiper}
P_{0}^{2}-P_{1}^{2}-P_{2}^{2}-P_{3}^{2}-P_{4}^{2}=-\frac{1}{l^{2}}
\end{equation}
and provide the four-dimensional de Sitter space which can be parametrized
by $p_{\mu }$.

From this point of view the space of momenta is not a flat space, as in
special relativity, but it is curved, maximally symmetric space of constant
curvature (this fact was already used within the DSR framework 
, see e.g. \cite{dsr}).

One can show that the general realization (\ref{gen_real_P}%
) for the $\kappa $ -Poincare momenta describe a momentum space with the
'generalized de Sitter metric' which leads to the 'relative-locality' effect as well:
\begin{eqnarray}
ds^{2} &=&\left[ -\frac{1}{l^{2}}\left( Z^{-1}\right) ^{\prime }Z^{\prime
}+\left( \left( Z^{-1}\right) ^{\prime }\left( \frac{1}{Z\varphi ^{2}}%
\right) ^{\prime }-\left( \frac{1}{Z\varphi }\right) ^{\prime 2}\right)
p_{i}^{2}\right] dp_{0}^{2} \\ \nonumber
&&-\left( \frac{1}{Z\varphi }\right) ^{2}dp_{i}^{2}+2\left( \left(
Z^{-1}\right) ^{\prime }Z^{-1}\varphi ^{-2}-\left( \frac{1}{Z\varphi }%
\right) ^{\prime }\frac{1}{Z\varphi }\right) p_{i}dp_{0}dp_{i},
\end{eqnarray}
where $\left( \cdot \right) ^{\prime }~ $ stands for $~\frac{d}{dp_{0}}$.
In fact the line element $ds^2$ above is a local expression for an induced metric on the hyperboloid (\ref{hiper}) written in local coordinate system provided by the formulas (\ref{gen_real_P}).

However to obtain the relative-locality effect (in the more general "framework" than in \cite{rel-loc3}), it is enough to consider the simpler cases, with the choice $
\psi =1$ for which the shift operator is $Z=e^{-lp_{0}}=e^{A}$, hence the
realization of momenta reduces to:
\begin{eqnarray}\label{gen_real_PZ}
P_{0}\left( p_{0},p_{i}\right) &=&\frac{1}{l}\sinh \left( lp_{0}\right) +%
\frac{lp_{i}^{2}}{2\varphi ^{2}}e^{lp_{0}};\\
 P_{i}\left(
p_{0},p_{i}\right)&=&\frac{p_{i}}{\varphi }e^{lp_{0}};
 \\
P_{4}\left( p_{0},p_{i}\right) &=&\frac{1}{l}\cosh \left( lp_{0}\right) -%
\frac{lp_{i}^{2}}{2\varphi ^{2}}e^{lp_{0}}.
\end{eqnarray}
For this choice $\varphi=\varphi_\lambda =Z^{-\lambda }=e^{-\lambda A}=e^{\lambda lp_{0}}$ ($\lambda$ is real).
Within this realization one gets the line element which depends on the parameter $\lambda$ and has the form:
\begin{equation}\label{metric}
ds_\lambda^{2}=\left[ 1-l^{2}\lambda ^{2}p_{i}^{2}e^{2\left( 1-\lambda \right)
lp_{0}}\right] dp_{0}^{2}-e^{2\left( 1-\lambda \right)
lp_{0}}dp_{i}^{2}+2l\lambda e^{2\left( 1-\lambda \right)
lp_{0}}p_{i}dp_{0}dp_{i}. 
\end{equation}

One can easily notice that for the choice of $\lambda =0$ we recover the Majid-Ruegg case\footnote{The convention in this letter differs from the one introduced in \cite{rel-loc3} by $l\rightarrow -l$.}: $
ds^{2}=dp_{0}^{2}-e^{2lp_{0}}dp_{i}^{2}$ with the so-called 'Majid-Ruegg metric' 
$g_{\mu \nu }=diag\left( 1,-e^{2lp_{0}},-e^{2lp_{0}},-e^{2lp_{0}}\right) $ 
\cite{rel-loc3}.

\section{Momentum space geodesics}
\subsection{Christoffel symbols}

From any metric one can calculate the Christoffel symbols from the general formula:
\begin{equation}
\Gamma_{\rho }^{\mu \nu }=\frac{1}{2}g_{\sigma \rho }\left( g^{\sigma \mu ,\nu
}+g^{\nu \sigma ,\mu }-g^{\mu \nu ,\sigma }\right).  \label{Chris}
\end{equation}

Limiting ourselves to the case of $\psi =1,$ $\varphi =Z^{-\lambda }=e^{-\lambda A}$, the nonzero components of the metric (\ref{metric}) are:
\begin{equation}
g_{00}=1-l^{2}\lambda ^{2}p_{i}^{2}e^{2\left( 1-\lambda \right) lp_{0}};\qquad g_{ki}=-e^{2\left( 1-\lambda \right) lp_{0}}\delta _{ki};\qquad
g_{0i}=g_{i0}=l\lambda e^{2\left( 1-\lambda \right) lp_{0}}p_{i}
.\end{equation}

The inverse metric is:
\begin{equation}
g^{\rho \sigma }={
\begin{pmatrix}
1 & l\lambda p_{1} & l\lambda p_{2} & l\lambda p_{3} \\[10pt] 
l\lambda p_{1} & l^{2}\lambda ^{2}p_{1}^{2}-e^{-2\left( 1-\lambda \right)
lp_{0}} & l^{2}\lambda ^{2}p_{1}p_{2} & l^{2}\lambda ^{2}p_{1}p_{3} \\ 
l\lambda p_{2} & l^{2}\lambda ^{2}p_{1}p_{2} & l^{2}\lambda
^{2}p_{2}^{2}-e^{-2\left( 1-\lambda \right) lp_{0}} & l^{2}\lambda
^{2}p_{2}p_{3} \\ 
l\lambda p_{3} & l^{2}\lambda ^{2}p_{1}p_{3} & l^{2}\lambda ^{2}p_{2}p_{3} & 
l^{2}\lambda ^{2}p_{3}^{2}-e^{-2\left( 1-\lambda \right) lp_{0}}%
\end{pmatrix}
}\end{equation}

For this choice of realization in the metric we obtain the following set of Christoffel symbols:
\begin{equation}
\Gamma_{i}^{0j}=-\left( 1-\lambda \right) l\delta _{i}^{j}=\Gamma_{i}^{j0};\qquad
\Gamma_{0}^{ij}=l\left( \lambda -(1-\lambda )\left( e^{-2l(1-\lambda
)p_{0}}-l^{2}\lambda ^{2}p_{i}^{2}\right) \right) \delta ^{ij};
\end{equation}
\begin{equation}
\Gamma_{0}^{i0}=l^{2}(1-\lambda )\lambda p^{i}=\Gamma_{0}^{0i};\qquad 
 \Gamma_{k}^{ij}=-l^{2}(1-\lambda )\lambda p_{k}\delta ^{ij};
\end{equation}
\begin{equation}
\Gamma_{0}^{00}=0;\qquad \Gamma_{k}^{00}=0.
\end{equation}
It can be seen that, within the first order in deformation, the
components $\Gamma_{0}^{0j}$ and $\Gamma_{k}^{ij}$ vanish
\begin{equation}
\Gamma_{0}^{0j} = {\mathcal{O}}(l^{2}); 
\qquad  \Gamma_{k}^{ij} = {\mathcal{O}}(l^{2}).
\end{equation}

For the sake of comparison with the results in Ref.\cite{rel-loc3}, we
give the explicit expressions of the above quantities for the special case
of $\lambda =0$: 
\begin{equation}
  \Gamma_{i}^{0j} =\Gamma_{i}^{j0}=-l\delta
_{ij};  \qquad {%
   \Gamma_{0}^{ij}=-}le^{-2lp_{0}}\delta ^{ij};
\end{equation}
\begin{equation}
\Gamma_{0}^{0j}=0; \qquad 
\Gamma_{k}^{ij} = 0.
\end{equation}

\subsection{Geodesic equation}
In this chapter and later on our focus is directed  only to the first order in the deformation parameter $l$. 
The geodesic equation in momentum space reads as:
\begin{equation}
\ddot{p}_{\rho }+\Gamma_{\rho }^{\mu \nu }\dot{p}_{\mu }\dot{p}_{\nu }=0,
\end{equation}
where $\dot{\ } ~ $ stands for $~\frac{d}{ds}$ and $s$ denotes a geodesic parametrization.

For the solution of the geodesic equation up to the first order in the deformation
parameter $l$ we can use the following ansatz \cite{rel-loc3}
\begin{equation}
p_{\rho }\left( s\right) =P_{\rho }s+\frac{1}{2}\Gamma_{\rho }^{\mu \nu }P_{\mu
}P_{\nu }\left( s-s^{2}\right); \end{equation}

\begin{equation}
\dot{p}_{\rho }\left( s\right) =P_{\rho }+\frac{1}{2}\Gamma_{\rho }^{\mu \nu
}P_{\mu }P_{\nu }\left( 1-2s\right), \end{equation}
with the initial conditions: $p_{\mu }\left( 0\right) =0; p_{\mu }\left( 1\right) =P_{\mu }$.

Also the inverse metric in the linear order in $l$ has the easier form
\begin{equation}
g^{\rho \sigma } ={ 
\begin{pmatrix}
1 & l\lambda p_{1} & l\lambda p_{2} & l\lambda p_{3} \\[10pt] 
l\lambda p_{1} & -1+2\left( 1-\lambda \right) lp_{0} + {\mathcal{O}}(l^{2}) & 0 & 0 \\ 
l\lambda p_{2} & 0 & -1+2\left( 1-\lambda \right) lp_{0} + {\mathcal{O}}(l^{2}) & 0 \\ 
l\lambda p_{3} & 0 & 0 & -1+2\left( 1-\lambda \right) lp_{0} + {\mathcal{O}}(l^{2})%
\end{pmatrix}
}
\end{equation}
There are only two non-zero Christoffel symbols in this case:
\begin{equation}
\Gamma_{i}^{0j}=-\left( 1-\lambda \right) l\delta _{i}^{j};\qquad
\Gamma_{0}^{ij}=l\left( 2\lambda -1\right) \delta ^{ij}.
\end{equation}
Therefore our solutions read as follows:\\
$p_{0}\left( s\right) =P_{0}s+\frac{l}{2}\left( 2\lambda -1\right)
P_{i}^{2}\left( s-s^{2}\right)\qquad \mbox{with}\qquad \dot{p}_{0}\left( s\right) =P_{0}+\frac{l}{2}\left( 2\lambda -1\right)
P_{i}^{2}\left( 1-2s\right)$\\
and $p_{i}\left( s\right) =P_{i}s-\left( 1-\lambda \right) l\delta
_{i}^{j}P_{0}P_{j}\left( s-s^{2}\right)\qquad \mbox{with}\qquad \dot{p}_{i}\left( s\right) =P_{i}-\left( 1-\lambda \right) l\delta
_{i}^{j}P_{0}P_{j}\left( 1-2s\right) $.

With this, it is straightforward to calculate the quadratic expression
$g^{\mu \nu }\dot{p}_{\mu }\left( s\right) \dot{p}_{\nu }\left( s\right)
= P_{0}^{2}-P_{i}^{2}+lP_{0}P_{i}^{2}+ {\mathcal{O}}(l^{2}),$  giving rise to the length of the momentum
space worldline.
Indeed, the length of the worldline,
$D\left( 0,P_{\mu }\right) =\int_{0}^{1}ds\sqrt{g^{\mu \nu }%
\dot{p}_{\mu }\left( s\right) \dot{p}_{\nu }\left( s\right) },$ 
 in momentum space between the two boundary
points, specified by the two values of the parameter $s,$ namely  $0$ and $1$ respectively, 
can be calculated within the first order in deformation $l$ as
\begin{equation}
D\left( 0,P_{\mu }\right) =\int_{0}^{1}ds\sqrt{P_{0}^{2}-P_{i}^{2}+lP_{0}P_{i}^{2}}=\sqrt{%
P_{0}^{2}-P_{i}^{2}+lP_{0}P_{i}^{2}}.
\end{equation}
Postulating that the geodesic distance from the origin to a generic point in
momentum space is the mass of a particle \cite{rel-loc1}, we get the relation:
\begin{equation}\label{m2}
m^{2}= P_{0}^{2}-P_{i}^{2}+lP_{0}P_{i}^{2}+ {\mathcal{O}}(l^{2}).
\end{equation}
The obtained result is the same as in \cite{rel-loc3}, therefore it is
realization independent, i.e. there is no explicit dependence on
$\lambda$. Since the mass Casimir should depend neither on the choice of the ordering nor realization,
 the results of the foregoing calculations show that the above
 postulate makes sense and is thus
  physically reasonable (relation between ordering and realizations is discussed in \cite{mssg}). Nevertheless, it seems that the
physical phenomena, as the time delay, will depend on realization for the
noncommutative coordinates, at least within the class of realizations
considered in this paper, parametrized by the parameter $\lambda $. And this point
will be shown in the next chapter.

\section{Hamiltonian description and time delay}
The momenta realization introduced above corresponds to a certain
realization of noncommutative ($\kappa$-Minkowski) spacetime coordinates: 
\begin{equation}  \label{real}
\hat{x}_{0}=x_{0}\psi \left( A\right) -lx_{k}p_{k}\gamma \left( A\right)
,\;\;\;\hat{x}_{i}=x_{i}\varphi \left( A\right)
\end{equation}
for an arbitrary choice of $\psi $, $\varphi $, where $\varphi$ is the same function appearing in the momentum realization (\ref{gen_real_P}). These functions satisfy: $%
\gamma =\frac{\varphi ^{\prime }}{\varphi }\psi +1$ with the initial
conditions: $\psi (0)=\varphi \left( 0\right) =1,\varphi ^{\prime }\left(
0\right) $-finite and $A=ia\cdot \partial =-a\cdot p$. (with $a=(l,0,0,0)$ as before) with $\varphi^{\prime }=\frac{\partial \varphi }{\partial A}$. A special case of the above, when one chooses: $\varphi_\lambda =Z^{-\lambda };\psi
=1;\gamma =\left( 1-\lambda \right) $ and 
\begin{equation}
\hat{x}_{0}=x_{0}-l\left( 1-\lambda \right) x_{k}p_{k},\;\;\;\hat{x}%
_{i}=x_{i}Z^{-\lambda }  \label{realZ}
\end{equation}
will be used in the calculations below.

Such realizations (\ref{real},\ref{realZ}) satisfy the following ($\kappa$-Minkowski) commutation relations:
\begin{equation}
\left[ \hat{x}_{0},\hat{x}_{i}\right] =il\hat{x}_{i};\qquad [\hat{x}_i,\hat{x}_k]=0.
\end{equation}
$\kappa $ -deformed phase space with deformed Poisson brackets can be
obtained by the so-called "dequantization" procedure: $\{\quad ,\quad \}=\frac{1}{i}\left[ \quad ,\quad \right] $. In this way we obtain:
\begin{equation}
\{x_{0},x_{i}\}=lx_{i}; \qquad \{x_{i},x_{j}\}=0,
\end{equation} together with
\begin{eqnarray} \label{pb21}
\{p_{0},x_{0}\}=1;\qquad\{p_{0},x_{i}\}=0;\\
\{p_{i},x_{0}\}=l\left( 1-\lambda \right) p_{i};\qquad
\{p_{i},x_{j}\}=-e^{\lambda lp_{0}}\delta _{ij}. \label{pb22}
\end{eqnarray}
It is easy to see that
the realizations  (\ref{realZ}) in conjunction with the ordinary Heisenberg algebra $\left[ p_{\mu },x_{\nu
}\right] =i\eta _{\mu \nu }$ lead to a phase space commutation
relations, which through the above described dequantisation procedure
come up with the momentum space Poisson brackets (\ref{pb21}) and (\ref{pb22}).

The previously obtained linearized relation $%
m^{2}=p_{0}^{2}-p_{i}^{2}+lp_{0}p_{i}^{2}\,$ can be used to postulate the form of the
Hamiltonian \cite{ham1,ham2} as:
\begin{equation}\mathcal{H}=\mathcal{N}\left( p_{0}^{2}-p_{i}^{2}+lp_{0}p_i^2-m^{2}\right),\end{equation}
where $\mathcal{N}$ is the constant multiplier.
Even though the on-shell relation  (\ref{m2}) does
not depend on the realization, the parameter $\lambda $ will enter the
particle's velocity and worldline through the Poisson brackets (\ref{pb22}).
This is made obvious by writing down the
Hamilton equations for the particle coordinates, which give rise to \footnote{For simplicity we consider 1+1 dim case.}:
\begin{eqnarray}
\dot{x}_{0}=-\mathcal{N}\left( 2p_{0}+lp_{1}^{2}+\left( 2lp_{0}p_{1}-2p_{1}\right)
l\left( 1-\lambda \right) p_{1}\right) ;\\
\dot{x}_{1}=-2\mathcal{N}\left( lp_{0}p_{1}-p_{1}\right) e^{\lambda lp_{0}},
\end{eqnarray}
with the corresponding equations for the particle momenta being trivial.
This leads to the velocity of a particle (in general):
\begin{equation}
v=\frac{2\left( lp_{0}p_{1}-p_{1}\right) e^{\lambda lp_{0}}}{%
2p_{0}+lp_{1}^{2}+\left( 2lp_{0}p_{1}-2p_{1}\right) l\left( 1-\lambda
\right) p_{1}}
\end{equation}
and in the leading order in $l$:
\begin{equation}v = -\frac{p_{1}}{\sqrt{m^{2}+p_{1}^{2}}}-\left(
\lambda -1\right) lp_{1}\frac{m^{2}}{m^{2}+p_{1}^{2}} + {\mathcal{O}}(l^{2}).
\end{equation}
Therefore the worldline of the particle appears to be given by
\begin{equation}
x^{1}=\bar{x}^1+v\left( x^{0}-\bar{x}^{0}\right) =\bar{x}^{1}-\left( 
\frac{p_{1}}{\sqrt{m^{2}+p_{1}^{2}}}+\left( \lambda -1\right) lp_{1}\frac{%
m^{2}}{m^{2}+p_{1}^{2}}\right) \left( x^{0}-\bar{x}^{0}\right),
\end{equation}
where $\bar{x}^0,\bar{x}^{1}$ are the initial time and position, respectively.

One can notice that the worldline for the massless particle is momentum and
realization independent:
\begin{equation}x^{1}=\bar{x}^{1}-\frac{p_{1}}{\left\vert p_{1}\right\vert }\left( x^{0}-%
\bar{x}^{0}\right).
\end{equation}

However this fact does not imply that simultaneous emission of such particles
with different momenta will be detected simultaneously \cite{rel-loc4}. This appears to be one of the properties of relative-locality idea.
Following the analogous analysis to the one performed in \cite{rel-loc3}, we
obtain the correction to the difference of Bob's detection times  for the
two particles sent by Alice:
\begin{equation}
\Delta t=lb\left( 1-\lambda \right) \Delta p_{1},\end{equation}
where  $b$ is the distance between Alice and Bob and $\Delta p_{1}$ is
the momentum difference between two photons emitted from the position of Alice (cf. \cite{us}). It is evident from the analysis (see also \cite{rel-loc3}) that the two  events, each of which corresponding to a single photon being  
registered by a detector, appear differently to two mutually remote  
observers. While  for one observer (Alice) these two events  appear as  
simultaneous, for the other observer (Bob) they do not occur  
simultaneously. This kind of peculiarity is a characteristic of  
relative-locality. In a case that the two observers are close to each  
other (in which  case $b$ is small), the product $lb$ will practically  
vanish due to $l$ being of the order of the Planck length, and the  
effect will not show up. On the contrary, if the two observers are far  
away from each other (in which case $b$ tends to infinity), the effect  
is more likely to occur. Thus, greater the distance between two  
observers, more tangible the relative-locality effect will be \cite{rel-loc1},\cite{rel-loc2},\cite{rel-loc3},\cite{rel-loc4}.
    The origin of this feature can be sought in a peculiar geometry of  
the phase space, which particularly comes into prominence when the two  
observers need to communicate and share among themselves their own  
descriptions of the same physical events.

One can notice that for $\lambda$=0 (right-ordering) we recover the
result from \cite{rel-loc3}, while for $\lambda$=1, the case which corresponds to the left-ordering, there is no Planck scale effect at all.

\section{Conclusion} 

In this Letter we have considered a large class of realizations of the
momentum sector of $\kappa$-Poincare algebra and have studied the
effect of the variation of realizations on the expressions for the
mass as well as the time delay formulae as obtained within the DSR
framework. The mass formula obtained in \cite{rel-loc3} using the
Majid-Ruegg bicrossproduct realization agrees with that obtained in
this Letter. This indicates the existence of a universality in the
mass formula for a wide class of realizations. On the other hand, the
time delay formula clearly shows a dependence on the choice of
realization. This is interesting from a phenomenological point of
view, since observations of time delays of signals coming from a GRB
can be used to put constraint on the allowed class of realizations. 

 Here we come to the main results of our paper. The relative-locality framework, with its curved momentum space geometry and
    nontrivial symplectic phase space structure leads to physical
    features that challenge our basic notions of spacetime locality.
  This framework leads to phenomena which exhibit a relative-locality, a notorious example of which is the presence
    of time delay in detecting of two simultaneously emitted
    photons. More precisely, while the observer at the emitter will
    see two simultaneously emitted photons as arriving at the detector
    with no time separation, the other observer, located at the
    detector will see the same two simultaneously emitted photons
    as coming at the detector with some time delay. What we found is
    that this time difference in two photons reaching the detector, as
    observed by the observer located at the detector, is
    realization dependent. Moreover, while the massive particles
    appear to have velocities that are realization dependent, the massless
    particles such as photons have velocities that are realization independent. 

 A particular choice of the ordering prescription may also appear to
  be important in other physical contexts, such as that of quantum
  statistics. This was demonstrated to be the case by mutual
  comparison of the oscillator algebras obtained in a number of
   different works
 \cite{Daszkiewicz:2007az},\cite{Arzano:2007ef},\cite{Daszkiewicz:2007ru},\cite{Young:2007ag},\cite{gghmm},\cite{Young:2008zg},\cite{gghmm2}.
    However, from this
  perspective, it is quite interesting to note that for a class
   of orderings/realizations of the $\kappa$-Minkowski space considered in
  this paper, there exists a 
   universal $R$-matrix, the same for all realizations within this class,  leading to the same algebra of creation
  and annihilation operators appearing in the mode expansion of the
  field operator and consequently leading to the same particle statistics.
What would be even more intriguing is to have this $R$-matrix fully
  expressed in terms of the Poincare generators, which would provide
  a unique covariant definition of the particle exchange, as well
  as the covariant notion of identical particles in the $\kappa$-deformed field theories. Some
  progress in this direction has been done in the triangular
  quasibialgebra setting of Ref.\cite{Young:2008zm}  and in the
  $\kappa$-deformed phase space approach related to a bialgebroid structure \cite{Meljanac:2012fa}.
Another issue is the choice of the metric on the deformed momentum
  space. Within
 the introduced framework, it would be interesting to investigate
  whether, e.g. the momentum space metric introduced via the
  commutation relations for the deformed Lorentz generators
  \cite{kmps} would also lead to the similar relative-locality
  effects. In the same context it would also be interesting to see what
  would be the mass relation calculated via geodesic distance.

\section*{Acknowledgements}
We would like to thank A. Borowiec for helpful remarks and discussion.
 S.M. and A.S work was supported by the Ministry of Science and Technology of the Republic of Croatia under Contract No. 098-0000000-2865. A.P. acknowledges the financial support of Polish NCN Grant No. 2011/01/B/ST2/03354.\\
\textbf{Note added}: 
We were informed by G. Amelino-Camelia that he and his group are analyzing similar issues.

\end{document}